\documentclass[sigconf]{acmart}
\AtBeginDocument{%
  \providecommand\BibTeX{{%
    \normalfont B\kern-0.5em{\scshape i\kern-0.25em b}\kern-0.8em\TeX}}}

\setcopyright{none}
\renewcommand\footnotetextcopyrightpermission[1]{}
\usepackage{multirow}
\usepackage{array}

\usepackage{balance}       
\usepackage{graphics}      
\usepackage[T1]{fontenc}   
\usepackage{mathptmx}
\usepackage{color}
\usepackage{booktabs}
\usepackage{textcomp}
\usepackage{makecell}
\usepackage{xspace}
\usepackage{fancyvrb}
\usepackage{amsthm}
\usepackage{hyperref}

\usepackage{microtype}        
\usepackage{ccicons}          

\newcommand{\eg}{\textit{e.g.}\@\xspace}

\begin{document}

\title{Privacy by Design: Bringing User Awareness to Privacy Risks in Internet of Things}

\author{Usama Younus}
\authornote{Both authors contributed equally to this research.}
\email{uyounus@umd.edu}
\affiliation{%
  \institution{University of Maryland}
  \city{College Park}
  \state{Maryland}
  \country{USA}
  \postcode{20740}
}

\author{Rie Kamikubo}
\authornotemark[1]
\email{rkamikub@umd.edu}
\affiliation{%
  \institution{University of Maryland}
  \city{College Park}
  \state{Maryland}
  \country{USA}}

\renewcommand{\shortauthors}{Younus and Kamikubo}
\begin{abstract}
This paper aims to cover and summarize the field of IoT and related privacy concerns through the lens of privacy-by-design. With the ever-increasing incorporation of technology within our daily lives and an ever-growing active research into smart devices and technologies, the concerns for privacy are inevitable. We intend to briefly cover the broad topic of privacy in the IoT space, the inherent challenges and risks in such systems, a few recent techniques that intend to resolve these issues on sub-domain level and on a system scale level. We then proceed to approach this situation through design thinking and privacy-by-design, given that most of the prior efforts are based on resolving privacy concerns on technical grounds with system level design. We participated in co-design workshop for privacy of a content creation platform and used those findings to deploy a survey based mechanism to tackle some key concern areas for user groups and formulate design principles for privacy that promote transparent, user-centered, and awareness provoking privacy design. 
\end{abstract}


\keywords{Internet of Things, Privacy Awareness, Risks}


\maketitle
\pagestyle{plain}

\section{Introduction}

The Internet of Things, in itself, covers a broad spectrum of topics and devices. The field emerged from a number of overlapping trends ranging from cheap sensors, high computing powers, miniaturization of design, accurate location positioning systems to increased and enhanced and inexpensive network access, and widespread use or mobile phones as interface devices. The devices included in a connected IoT space do not correspond to multi-use computation resources such as phones, tablets or laptops. Instead, they refer to products built for a narrow range of utility as listed in Table~\ref{tab:IoT_examples} and they have the ability to articulate one or all of the three behaviours: sense, analyze and communicate.

To fully leverage the user experiences of IoT for wider applications, it is important to understand privacy expectations that people have and ensure that the technologies meet individual privacy requirements. How IoT devices are collecting and using a broad range of user activities and behaviors, often silently in the background, has raised one of the major concerns in the IoT world. Naeini et al.~\cite{naeini2017privacy} conducted a survey study to identify the contributing factors in data collection and use that impact individual privacy preferences and comfort levels. The invasive nature of collecting data in private places such as homes and acquiring sensitive personal information such as biometrics had a negative impact on the perceived comfort, and the survey respondents preferred to be notified about data practices, such as sharing of data to third parties or data retention period. However, we are yet limited in understanding how we can support their privacy goals and expectations while navigating the risks and benefits associated with data practices in the IoT space. User awareness of such tensions and approaches to navigate around them are not discussed in bringing design implications for future IoT devices and applications.

\begin{table}
\centering
\begin{tabular}{>{\centering}p{0.4\linewidth}>{\centering\arraybackslash}p{0.53\linewidth}}
\multicolumn{2}{c}{\textbf{IoT Applications and Devices}}\\\cline{1-2}Consumer&Enterprise\\
\hline
\begin{itemize}
  \item smart speakers
  \item connected cars
  \item wearable fitness trackers
  \item smart lighting
  \item smart thermostats
  \item smart TVs
  \item robot vacuums
  \item smart locks
  \end{itemize}
  &
\begin{itemize}
  \item automated retail checkout
  \item face recognition cameras for security
  \item predictive equipment maintenance 
  \item temperature-sensitive supply chain
  \item employee wellness trackers
  \item drones
  \end{itemize} \\
\hline
\end{tabular}
  \caption{Consumer \& Enterprise based IoT Examples. \cite{rosner2018privacy}}
  \Description{A number of IoT examples in the domestic and industrial environment }
  \label{tab:IoT_examples}
\end{table}

In this paper, we address open challenges in the space of IoT to establish a holistic framework for privacy guarantees that amplify user awareness of privacy risks and allow user management and control towards privacy. While asking the user to agree to the terms of service agreement and privacy policy is a minimal effort that companies can make to bring user awareness to privacy implications, it is not the best practice as the terms are often signed without being read~\cite{guynn2020what}. Given the lack of user awareness and control towards privacy, there is a significant importance placed on techniques and algorithms with privacy guarantees implemented at different levels of system entities in the IoT (i.e., in the device, infrastructure, and service layers). However, privacy needs and preferences of people subject to data collection and receiving services are often outside of this conversation.

Discussions for greater user awareness and control towards privacy in the IoT world are not new~\cite{mehrotra2016tippers}. Privacy studies in different research domains, including information technology and HCI, have called for better notifications and consent procedures on IoT devices~\cite{rosner2018privacy,schaub2015design}. Our work is complementary to but qualitatively different from these prior efforts that include surveying the existing literature on privacy notices to develop a design space for contextually relevant notifications~\cite{schaub2015design}. Instead, we aim to gain a deeper understanding of privacy goals and expectations of privacy risks that people have that would improve the existing efforts to be more complete, realistic, and user centered. To achieve this goal, we employ a design approach i.e. Privacy by Design by conducting a survey with 16 participants to guide design of a privacy inclusive solution for IoT applications and devices. Based on the survey findings, we aim to discuss implications for developers to encourage UI and UX of the IoT space to have more privacy centered design.

\section{Internet of Things 101}
Starting with the vision of ubiquitous computing \cite{weiser1991computing}, the IoT depends on technological progress to bring increasing miniaturization and availability of information and communication technology at decreasing cost and energy-consumption. Here we present the privacy definition which we believe sufficiently covers the broader spectrum of the IoT. We then move onto privacy concerns from the perspectives of end users, and the key technologies related to these concerns in the past and their evolution. Finally, we evaluate the types of threats present in such systems.

\subsection{Privacy Definition \& IoT Reference Model} 

Considering the very broad and diverse spectrum of IoT devices, literature offers many definitions and perspectives related to privacy, with the \textit{information privacy} at the forefront of them all. It was defined by Westin in 1968 as "the right to select what personal information about me is known to what people". Although this definition, which referred to non-electronic devices, is still valid, it is at the same time too generic to enable any focused analysis of the topic. Hence, for the paper, we adopt a definition and reference model used in \cite{ziegeldorf2014privacy} for a more detailed approach:\newline

\textit{Privacy in the Internet of Things is the threefold guarantee to the subject for}
\begin{itemize}
    \item \textit{ awareness of privacy risks imposed by smart things and services surrounding the data subject }
    \item \textit{ individual control over the collection and processing of personal information by the surrounding smart things }
    \item \textit{ awareness and control of subsequent use and dissemination of personal information by those entities to any entity outside the subject’s personal control sphere }
\end{itemize}

\begin{figure}[h]
  \centering
  \includegraphics[width=\linewidth]{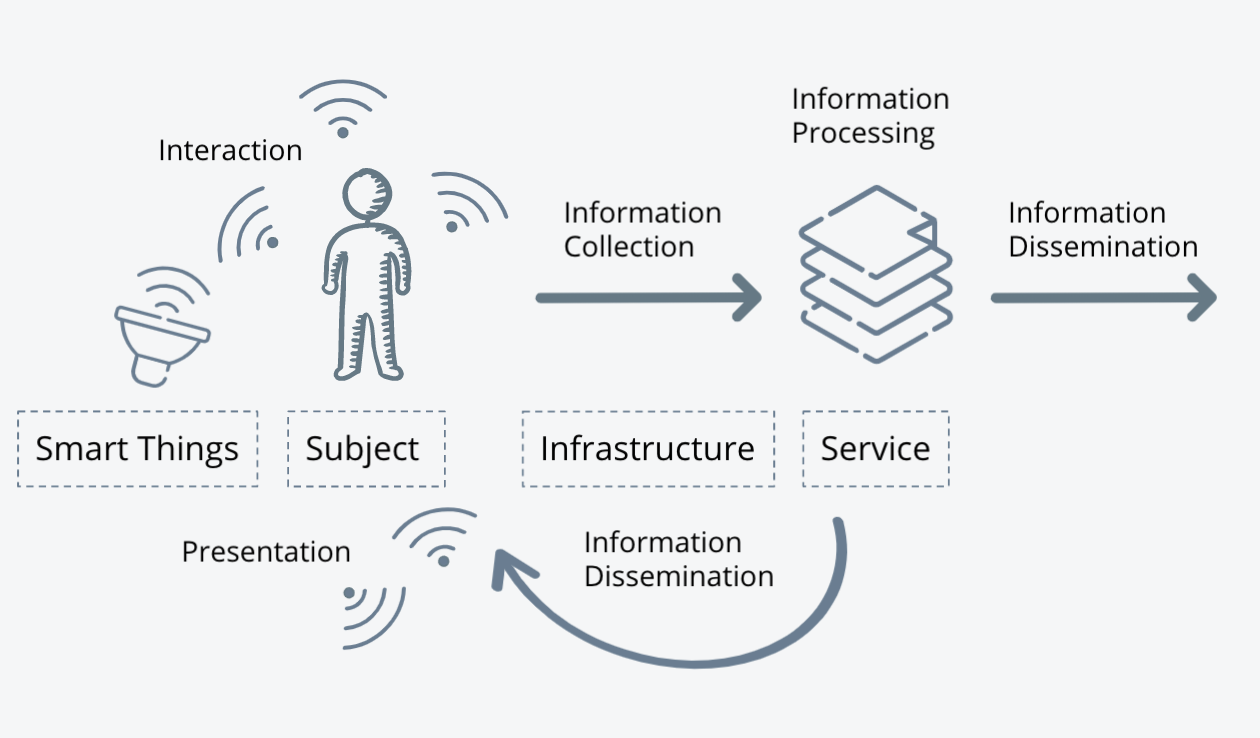}
  \caption{IoT Constitution: Sensing, analyzing and communicating involving smart things, subject to data collection who is also a recipient of service, and infrastructure connecting smart things and service. \cite{ziegeldorf2014privacy}}
  \label{fig:IoT_module}
\end{figure}

The IoT reference model that \cite{ziegeldorf2014privacy} refers to is based on the ITU \cite{union2005internet} and IERC \cite{vermesan2011internet} visions of the IoT and can be summarized as: \textit{Anyone and anything is interconnected anywhere at any time via any network participating in any service}. This reference model describes the entities and information flows of IoT applications.

In the model \textbf{four main types of entities} are under consideration. \textbf{\textit{Smart things}} are everyday things augmented with information and communication technology (ICT), with the ability to collect, process and communicate data about themselves and their environment with other things and humans. Backends host \textbf{\textit{services}} that gather, combine and analyze data from many smart things to offer a value-addition service to the end-user. Humans have two different roles here: either \textbf{\textit{subject}} to data collection by the smart things or \textbf{\textit{recipients}} of data or services. Finally, smart things are connected to services via an \textbf{\textit{infrastructure}} with different characteristics ranging from low-power lossy networks to powerful Internet backbones possibly traversing different intermediate gateways and servers, e.g. firewalls and protocol bridges.

This model abstracts most of the device specific details, interconnection methods and services, and hence provides a high-level overview of the wide range of IoT systems and applications. Though it might not hold true for every scenario, but it does convey a broader sense of the interconnections of IoT in our daily life.  

\subsection{Threats, Types of Attacks}
With the evolution of network-connected devices, offering opportunities for customers and businesses, we cannot omit the discussions on privacy risks~\cite{federal2015internet,rosner2018privacy,ziegeldorf2014privacy}. The technologies and features involved, as well as the new ways of interaction with the IoT, pose specific threats and attacks that need to be addressed~\cite{ziegeldorf2014privacy}.

\begin{description}
\item[Identification] The threat of identification lies in associating individuals with personal data, such as name or address. It has received major attention in the field of privacy, as collection of personal information is often concentrated and analyzed in a location outside of the end user's control~\cite{ziegeldorf2014privacy}. The threat is amplified with the nature of IoT devices adopted in private spaces. It diminishes the user’s conscious awareness of personal identifiable information being collected and managed, and the unobtrusive presence of network-connected devices can cause `black box' effects of losing sense of control and privacy~\cite{rosner2018privacy}. 
\item[Localization and Tracking] The threat of determining and recording a person’s location has been a major concern in the phase of information processing. Smart systems leave data trails that allow to track user activities in the environment. These location traces could lead to privacy violations, such as GPS stalking~\cite{voelcker2006stalked} or up-to-the-second details about illness~\cite{chow2009privacy}. What's worse, users are often not aware of when they are being tracked~\cite{rosner2018privacy}, such as observed in the case of indoor location based systems~\cite{chow2009privacy}. This threat also leads to other issues like inventory attacks; for example, burglars can use inventory information to target break-ins at private homes~\cite{ziegeldorf2014privacy}
\item[Profiling] The threat of profiling also brings issues in a privacy violating context. With the massive amount of information being collected from the interconnected devices, often in the immediate physical proximity, they can be used to build models to infer about individuals. Profiling methods are commonly found in personalization of advertisements or recommender systems. Data that can be collected from previously inaccessible parts of people’s private spaces advance in its quality, and the extent of profiling in IoT can negatively impact the decisions placed on the individuals, such as location-based price discrimination or unsolicited advertisements~\cite{toch2012personalization}. Data management needs to be carefully protected from data-selling businesses. 
\item[Lifecycle transitions]
Privacy is threatened when many IoT devices have a limited life cycle, which results in a risk with unsupported devices vulnerable to critical security or privacy bugs~\cite{federal2015internet}. Companies, particularly those developing low-end devices, may decide to not manage ongoing support or software security updates~\cite{schneier2014internet}, leaving consumers with security risks after purchase. Privacy issues also rise in the context of shared use cases of IoT devices~\cite{ziegeldorf2014privacy}. While once many electronics were limited to a single owner, following the buy-once-own-forever model, they now feature a much more dynamic lifecycle. People sell or dispose devices at a frequent pace to exchange with newer models or items. Without taking an appropriate action by the users and businesses that go beyond a memory wipe, there is a chance for unexpected disclosure of collected and stored information~\cite{nbc2021women}.
\item[Linkage]Ubiquitous data collection allows communication with other devices and third parties. Connected devices pose problems concerning what third-party services are allowed to collect and process user data (e.g., Alexa smart speaker connecting to Apple Music and Podcast). The combination of data sources reveals information that people did not disclose to the original source, or worse, did not want to reveal~\cite{ziegeldorf2014privacy}. Moreover, linkage of data sources and systems creates an increased risk of re-identification. Even with the common approach to use anonymized data, the act of combining different sets of anonymous data can often result in unforeseen attacks by adversaries~\cite{narayanan2010myths, hassan2020differential}.


\end{description}
\section{Prior Privacy Efforts}
Through our earlier illustration of the exponential growth of IoT space, we can fairly make an assumption that a significant amount of effort and resources would be deployed for the mitigation of the risks and privacy concerns surrounding this space. In this section, we explore the two forked approaches towards resolving these issues: \textbf{Systems' Approach} \& \textbf{Design Approach}.

\subsection{Systems' Approach}

In this section of the survey, we attempt to approach the different entities, as portrayed in Figure \ref{fig:IoT_module}, in a modular manner. We tend to discuss and elaborate a few recent privacy approaches in each of these sub-modules, and how they are attempting to secure that specific sub-area. Towards the end of this section, we also put reference to a few system level design approaches that cover all these aspects in a singular approach.

\subsubsection{Data Collection}
The data collection in an IoT setting can vary from data collection by different sensors in an environment to collection (transmission) of data from different IoT devices to central servers. The approaches here tackle scenarios, where we may have want to preserve the privacy of individual devices or prevent personal identification in data collection  process. One such approach, as cited in paper \cite{bonawitz2017practical} uses aggregation of high-dimensional data for transfer purposes to the server. This approach uses \textbf{data vectors for translation of user-data} into communication friendly format, which can be later securely communicated or even used in federated learning setting without learning each user's individual contribution. Another approach, which is used by Apple.inc as well, is the one referred in paper \cite{erlingsson2014rappor}. This is a \textbf{crowd-sourcing technology} that is used to obtain \textit{statistics} from the end-user client software, is motivated from lack of randomization for strings and non-trusted data collectors. One such usage is the collection of app time of browser use on client computers. The entire process works by utilizing the hash maps and bloom filters to convert information into bit-maps, followed by randomization on each bit to provide a layered approach to privacy. Lastly, an approach that has received traction in the field of privacy is the use of \textbf{locally deferentially private} (LDP) algorithms. The guarantees provided by these algorithms provide a strong set of guarantees, however even these degrade if the collection process is performed regularly. The paper \cite{ding2017collecting} tackles this problem by developing LDP mechanisms geared towards collection of counter data with improved guarantees. This technique is being used by Microsoft for their app data collection.

\subsubsection{Data Storage \& Handling}
This section refers to mechanisms that are geared at providing additional security to the remote IoT devices themselves (i.e. the sensing module which stores the sensed data). The premise for this section originates from scenarios where we have a compromised data storing device. One paper \cite{yi2015privacy} specially narrows the topic to data leakages by an internal resource. As in cases of medical records, data is stored and maintained, at times, by doctors etc. So in order to prevent data leakage by storage or collector, the paper uses an en encryption and distributed storage mechanism over multiple storage devices/points for privacy and query management. Another paper \cite{dorri2017blockchain} presents a \textbf{block-chain based model} for data handling in a smart home system. Though block-chain based approaches provide decentralized security and privacy, they still involve a great overhead in terms of energy, delay, and computational requirements.

\subsubsection{Secure Communication}
The secure communication protocols deal with unsecured communication channels. The premise of research in this sub-domain lies on securing the communication routes, and ensuring no privacy leakage is occurring in the Sensor Networks. This threat can be present while transfer of data from devices to server, as well as during inter-device data management. A paper that presents a detailed analysis on \textbf{securing wireless sensor networks} and provides a suite of optimized security protocols for the purpose is \cite{perrig2002spins}, which builds on-top of SPINS and has two secure building blocks: \textbf{SNEP and µTESLA}. 
\begin{itemize}
    \item SNEP includes data confidentiality, two-party data authentication, and evidence of data freshness; and is a cryptographic protocol
    \item µTESLA provides authenticated broadcast for severely resource constrained environments in a asymmetric delayed disclosure of symmetric keys.
\end{itemize}

\subsubsection{System Scale Architectures}
Since the overall process of IoT constitutes sensing, communicating and computing, a few papers have made an attempt to design holistic approaches encompassing \textbf{end-to-end pipeline} of IoT. These papers present design constraints \& protocols, hardware requirements, techniques \& tools for privacy, and specific algorithms for fast processing of information for all the three layers of an IoT system, i.e. the device layer, the IoT infrastructure/platform layer, and the IoT application layer. This sort of integration ensures a smooth flow of entire paradigm, and guarantees complete privacy coverage. The paper \cite{jayaraman2017privacy} goes into providing an overall IoT privacy preserving architecture for ensuring privacy in an IoT computing environment. They divide the overall flow of the schema into: \textit{communication channel security mechanisms, authorisation \& access control, and privacy preservation}. The proposed architecture and proof of concept has been implemented and shown to work on extensions of OpenIoT - a widely used open source platform for IoT application development. Another article that covers the wide spectrum of the overall private IoT architecture is \cite{tawalbeh2020iot}. It explains in detail the broader genre of IoT, applications and the respective security concerns. The article covers the mechanism to design a secure and private IoT system hosted on current cloud providers (Amazon Web Services in their case). For this entire scheme, they introduced security protocols and critical management sessions  between each of the system layers to ensure the privacy of the users’ information. Security certificates were utilized to allow data transfer between the layers of the proposed cloud/edge enabled IoT model. This ensured reduced security vulnerability and protection from cyber-security threats.

\subsection{Design Approach}
Privacy and IoT have been explored across different research domains, including law and public policy~\cite{federal2015internet}, information technology~\cite{barth2006privacy}, or Human Computer Interaction~\cite{denning2014situ}. In this section, we provide a brief summary of approaches taking different angles from the approaches discussed in the previous section. They are centered around ``Privacy by Design'' methods that consider data practices on the front end by businesses and developers (e.g., developing policies that impose data minimization )~\cite{federal2015internet} and user interfaces to improve control on IoT devices (e.g., designing platforms that allow users to specify their privacy preferences)~\cite{mehrotra2016tippers}. 

We observed more design efforts that addressed the challenges to allow user control given the lack of information space on IoT devices to provide details on data collection and consent procedures. Denning et al. investigated how individuals perceive AR-style recording to discuss design considerations for privacy-mediating technologies and suggested mechanisms for notification, consent, and blocking~\cite{denning2014situ}. Pappachan et al. proposed a framework to design privacy-aware smart buildings to enforce privacy policies that meet privacy preferences when collecting user data or sharing it with building services~\cite{pappachan2017towards}. While discussions on privacy and IoT attracted many researchers in different disciplines, there still remain barriers to fully integrating meaningful privacy practices in the design and deployment of ubiquitous computing systems~\cite{lederer2004personal}.

\subsection{Remaining Privacy Concerns}

We have presented a number of risks and their scenarios, including but not limited to Identification, Localization and Tracking, Profiling, Lifecycle Transitions, Linkage. Though prior attempts from different disciplines, including security and cryptography, information technology, and HCI, have aimed to tackle and resolve each of these scenarios in an IoT framework, the lack of developer and user awareness on the topic itself can lead to tremendous losses. A holistic framework that lays out a design map for creating awareness for users in the real world, and poses the right questions for developing IoT applications with security and privacy, could promote more complete, realistic, and user centered solutions.
\section{Method}
In this section, we elaborate on a co-design workshop that we participated in and the successive study that was formulated by us based off the lessons and learning from the workshop. The design workshop was aimed at the privacy module of a research platform, where a group indulged in creating different personas and their requirements. The analysis from these was later explored and utilized to deploy a survey for better understanding of the privacy by design approach.

\subsection{Formative Study}

We participated in a 2-hour co-design session conducted over a Zoom call (with 8 other participants) on improving the social media platform that allows users to collaborate with each other online through sharing and storing of media content, and communicating about that content. The main focus of the session was to consider different ways to address permission and privacy goals for the interface of the proposed platform. Brainstorming session was performed with the aim of developing user personas, and mapping their critical interaction points with the platform. Specifically, based on each user persona (e.g. administrator, content moderator, content contributor), we discussed who can see/comment on what content; who can download what content; who can share what content; who can edit/delete what content; and who can make what content private.

\begin{figure}[h]
  \centering
  \includegraphics[width=\linewidth]{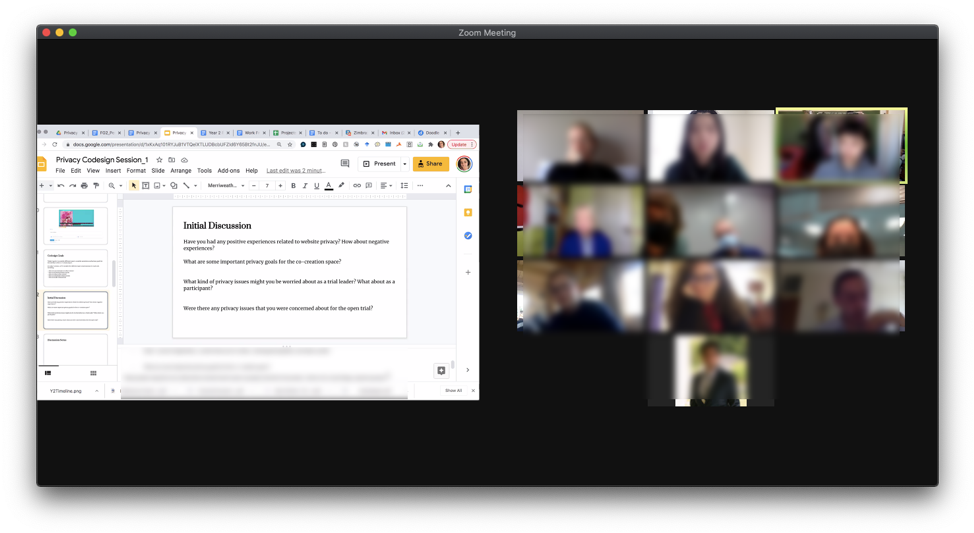}
  \caption{Privacy Co-Design Workshop. Acknowledgement: Alina Striner, Distributed and Interactive Systems Group}
  \Description{Acknowledgement: Alina Striner, Distributed and Interactive Systems Group}
  \label{fig:IoT_examples}
\end{figure}

The session further delved into the requirements of privacy in a general social network setting, and how each of the participants prior experiences have shaped their views of the matter. Through a shared understanding of the current content sharing venues, a few additional points of improvement were explored. Additionally, an effort was made by the organizer to connect the session with a number of previous sessions on platform design. This provided further insight into the overall schematics of the design, and what areas needed modification.

While we understand that the topic of privacy in this co-design session was not related to the IoT space, it involved a broad discussion on privacy goals, privacy issues, and personal experiences related to (social media) privacy. Our experiences and lessons learned from this session were utilized in designing our main study to similarly answer what personal information should be shared and how much information should be shared with different levels of entities in an IoT space.

\subsection{Survey Study}
We provide our reflective analysis and the design of our survey in this section.

\subsubsection{Study Design Motivation}
From our formative study, we were able to obtain some insights into how certain personas can and should manage the control over privacy, along with the need for levels (hierarchy) of control by user groups and preferences (\texttt{Privacy Notifications and Control}). Additionally through this, we were able to understand the need for anonymity (\texttt{Data Anonymity}), as well as informed usage of personal information  (\texttt{Privacy Agreement and Terms of Use}).

With hindsight, we saw that the formative study encompassed a broad range of views, from a very liberal view to conservative approach towards privacy, and how many of the associated threats inform user behaviour. One of the valid concerns raised was regarding the ownership of the data and monitoring it throughout the entire service pipelines (\texttt{Data Processing and Usage}). The concerns also extend to how the use of such information can be violated to infer about individuals that the service claims for understanding customer needs and improving user experience (\texttt{Data Profiling}). We believe all of these concerns can be addressed to some extent when approached through a user centered design and ease of access to relevant controls and preferences for data. 

\subsubsection{Survey Questions}

\begin{figure*}[h]
  \centering
  \includegraphics[width=0.8\linewidth]{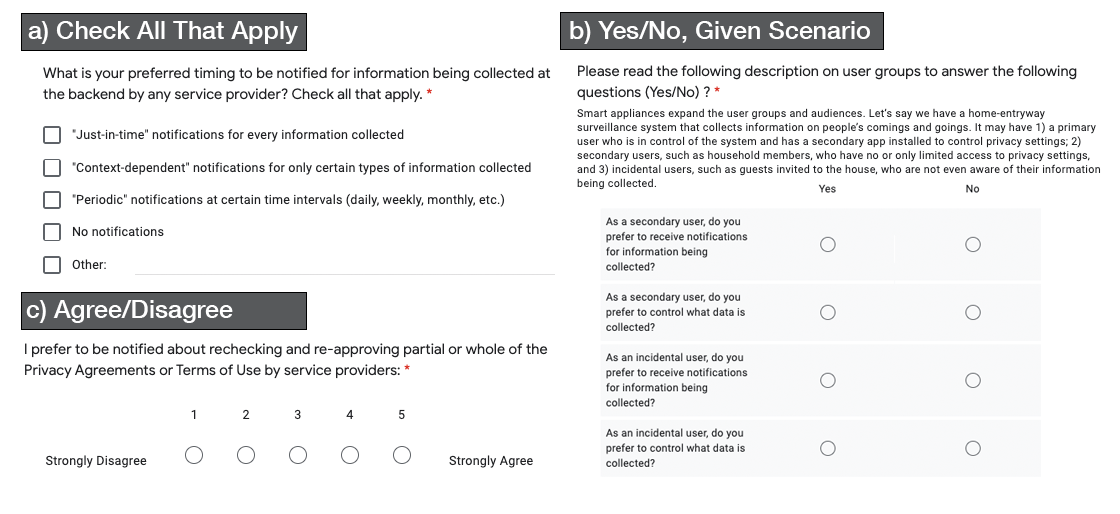}
  \caption{Example survey questions using Google Form designed in different formats: a)'check all that apply' with Other for open-ended comment option, b) 'dichotomous' yes or no according to a given scenario of user groups, c)'Likert scale' from 1 to 5, with 1 being strongly disagree and 5 being strongly agree to a given statement.}
  \label{fig:survey_examples}
\end{figure*}

The survey had 5 sections following the key themes emerged from the formative study: 
\begin{enumerate}
    \item \texttt{Privacy Notifications and Control} This section determines the design preference for notifications and control over data collected from a service provider, including the modality and timing to manage privacy implications. It has a list of design options where the respondents can select (\eg, See Figure~\ref{fig:survey_examples}a). We elaborate on privacy preference depending on different situations and user groups (\eg, See Figure~\ref{fig:survey_examples}b), with questions given in a dichotomous format (Yes/No) to make a clear distinction on their perspectives towards data practices. 
    \item \texttt{Privacy Agreement and Terms of Use} This section determines the user attitude towards privacy agreement and informed privacy decisions that they often need to make through click-to-agree contracts for privacy implications. We explained a background on how users tend to not read privacy policies, often leading to a lack of awareness of data practices. We dive into this problem by asking a series of 5-point Likert scale questions (\eg, See Figure~\ref{fig:survey_examples}c) to explore alternative ways to navigate through privacy agreements.
    \item \texttt{Data Anonymity} This section touches on the data processing technique that removes or modifies personally identifiable information and whether users are more or less supportive towards this technique. We again employ a 5-point Likert scale question format.
    \item \texttt{Data Processing and Usage} This section determines the design preference and improvement for data processing and management systems, such as deciding between local or remote/server data processing or considering having ownership tags linked with data collected if there is some compensation for it. To understand user perspectives, the questions are given in 5-point Likert scales, and there is a list of possible ownership tags that the respondents can choose from to match what they are comfortable with.
    \item \texttt{Data Profiling} Lastly, this section asks one 5-point scale question on the way data profiling results are elicited by service providers, whether respondents prefer to have full access and control on these results.
\end{enumerate}

\subsubsection{Data Collection} The overall survey study followed the procedure as summarized below:
\begin{description}
\item[Participant Recruitment] We collected 16 responses in total within a two-day period. We employed a non-probabilistic sampling method in the form of convenience sampling and word of mouth promotion. The link to the online survey was distributed to different groups in our school network through internal communication platforms (\eg Slack, Discord, GroupMe). The groups include graduate students and undergraduate students in different departments, including CS, Information Studies, Engineering, Math, and Psychology. No participation incentive was given. There was no inclusion criteria for participants, and we did not collect any of their personal information, such as name, gender, email address, or affiliated department \& organization.
\item[Ethical Approval]  Given our study design involving human subjects, we did not obtain research ethical approval. We followed instructions from our school’s Institutional Review Board division that the IRB approval is not required if the intent of our work is to meet course requirements, and results will not be used for something other than the course assignment. We therefore designed our survey for the purpose of the course assignment and did not obtain any personal information that would link the participants with the survey responses.
\end{description}

\section{Findings}

\subsection{Privacy Notifications and Control}

Design of notifications for privacy, such as visual prompts that notify the user how many times an app has used their location in the background, can impact users’ attention and awareness towards privacy, as well as their informed privacy decisions. These notifications can vary in the form of modality used; visual notifications, for example, can be provided as text, images, icons, or a combination thereof. As IoT devices are adopting multiple form factors to match the context of use such as a standalone smart speaker or a wearable fitness tracker, auditory or haptic feedback can also provide a potential modality to communicate privacy information. We investigated preferred notification design through our survey by allowing the participants to select or comment on possible modality options and user interface solutions for information feedback.

\subsubsection{Modality}
To understand the user perception towards such design of privacy notifications, we asked our participants about their preferred modality for notifications. The survey results showed that the majority of the participants selected ``visual’’ notifications about when data collection is occurring by a service provider, where 10 participants chose to receive them on a display screen of IoT devices and 9 participants chose to use a secondary app to receive notifications on their phone. In comparison, ``auditory'' and ``haptic'' notifications were selected by 3 participants each. 

We elaborated the notification design by having a set of options that the participants can choose: 1) showing a brief notification banner at the top of the screen that disappears in a few seconds and can be checked later in the list of notifications, 2) displaying a bell shaped icon for screen-based IoT devices, 3) displaying ring or button lights for auditory IoT devices, and 4) providing haptic vibration for wearable IoT devices. The results for visual notifications showed that bell-shaped icons received the highest preference score, selected by 10 participants, whereas ring or button lights were selected by 5 participants and notification banners were only selected by 1 participant. While we found that haptic notifications were not yet widely accepted for privacy notices compared to visual notifications, 5 participants selected haptic vibration as one of the design options. Although the participants were also given an option to choose `Other’ to share their ideas, we did not receive any response. This was expected as open-ended comments tend to get lower response rates, and we therefore provided a list of possible user interface solutions.

\subsubsection{Timing}
Timing of privacy notifications can have a significant impact on the effectiveness. Users’ attention and comprehension of privacy information can be negatively affected if the notifications are presented at an inopportune time (\eg, caused by distractions). We asked the participants to understand their perception of the most opportune time. We received responses from 10 participants agreeing with ``context-dependent’’ notifications to have the ability to choose certain types of data collection that they would get notified. The results also showed other possible timing opportunities including ``periodic’’ timing at certain intervals (daily, weekly, monthly, etc.) and ``just-in-time’’ for every information collected, which received 7 and 6 votes respectively. For ``Other,’’ we received preferences for privacy notifications shown at setup time before using IoT applications or devices or ahead of time before data are collected.

\subsubsection{Design for Different User Groups}
Lastly, we considered the best practices for privacy notice design in the IoT space by different user groups. A home-entryway surveillance system, for instance, collects information on people’s comings and goings, and it may have the following user groups: 1) a primary user who is in control of the system and has a secondary app installed to control privacy settings; 2) secondary users, such as household members, who have no or only limited access to privacy settings, and 3) incidental users, such as guests invited to the house, who are not even aware of their information being collected. The survey asked for yes/no responses to receive and control privacy implications given this scenario of the user groups. 

The survey results showed that in a total of 16 participants, 13 of them responded that as a secondary user, they prefer to receive notifications for information being collected and control what data is collected. We found a similar attitude towards receiving privacy notifications (14 yes responses) and controlling the data collected (12 yes responses) as an incidental user.

\subsection{Privacy Agreement}

Common regulations and business practices to inform consumers about privacy implications take the shape of privacy policies and terms of service. Service providers ask users to go through a user agreement at setup or install time so that they can inspect their data practices before using their service. However, in practice, the terms of service, privacy policies and other agreements are long, complex, and tend to be ignored and forgotten after the initial sign up~\cite{cranor2012necessary,mcdonald2008cost}. We address this problem by understanding user perspectives of and preferences for such agreements. 

\subsubsection{Navigating Terms of Service}
We asked our survey participants to see if there is a room for improvement in interacting with privacy agreements. While users have become largely habituated to ``I agree to’’ checkboxes only once at the beginning of the service use, participants responded that they prefer to be notified about rechecking and re-approving the terms of service throughout the use, with an average score of 4.31 (SD = 0.7) for their level of agreement to this preference in 5 points. We also investigated the right frequency for notifications for rechecking and re-approving. We received votes for quarterly (6), annual (5), bi-annual (4), weekly (2), and monthly (1). The open-ended comments also revealed 3 votes for when there is a change or update in the agreement. In addition, the majority of participants agreed that they prefer to have a privacy settings tutorial or run-through of privacy implications at the sign-up stage, with an average score of 3.94 (SD=1.12). 

\subsubsection{Making Initial Privacy Decisions}
We further explored this direction of user agreement as users are initially making informed privacy and consent decisions and going through the setup process. There is difficulty making informed decisions before the use of service when they cannot fully assess the utility and privacy tradeoffs. We asked our participants if they prefer to know about service degradation if there is a choice to opt out of collection of certain information. The survey results showed that the majority of participants agreed to this preference, with average scores of 4.06 (SD=1.34) and 4.31 (SD=1.01) representing their level of agreement to know the specific and general types of service, respectively, that will be degraded by opting out of certain data collection.

At setup, creating a profile account is often mandatory for users to start using the service. However, users may be focused on completing the setup process that they fail to pay attention to personal information they are giving out~\cite{good2007noticing}. We are interested in understanding participants’ level of agreement to different types of personal information that they are willing to share to the service provider. The survey results showed that they are most comfortable with sharing their email address, receiving the average score of 3.44 (SD=1.21). Other types of information were lower in the level of agreement, such as full name (M=2.88, SD=1.2), data of birth (M=1.94, SD=0.93), gender (M=2.38, SD=1.31), phone number (M=2.44, SD=1.15), home address (M=1.63, SD=1.02), although they are commonly asked during sign up for account authentication and password management or for interconnected services (\eg, Alexa voice shopping and delivery). Bank account information received the lowest level of agreement, with an average score of 1.06 (SD=0.25).

\subsection{Data Anonymity}
One of the oldest and widely used technique in data privacy is data anonymity. This refers to masking some to complete sets of private information for the purpose of making them anonymous and untraceable. Companies and data controllers rely on such techniques for maintaining the privacy of their large scale information flow - specifically this particular implementation which requires less expertise and computational resources (though this is not always guaranteed to provide fool-proof security \cite{ohm2009broken}). Despite the concerns around some failure incidents, this technique has found widespread acceptance globally on various institutional levels. Owing to its simplicity to understand, we tried to explore the user's perspective regarding this process.

Another reason for delving into this was our prior analysis which showed that the user's trust in system is enhanced as they get more intuition of their data practices. Since the sign-up process of any service requires giving-up some personal information that is usually kept in a secure manner, we wanted to evaluate if users will prefer to have control over this process of anonymity and security. This can further enhance the belief and comfort of users in using certain IoT services, as well as provide a means to service providers for justifying exactly which pieces of information are crucial to them.All questions in this section were measured over 5-point Likert scale.

\subsubsection{Anonymization Awareness}
Since the privacy expectations of users are somewhat driven by their awareness of the data practices of service providers, this question was particularly aimed at gauging whether the users will be interested in exploring which of their privately controlled information can be anonymized or not. Even though this step entails a firm belief in the privacy measures, but having a clear understating of the user awareness and requirement can lead to further refinement of the data collection process. The question specifically asked if the users want their service providers to share which of their private information can be anonymized. The users were to rate their preference on the 5-point Likert scale. The survey responses displayed overwhelming interest of users (M=4.62, SD=0.78) in obtaining this information by their providers, given a sustained quality of service is provided after this practice.

\subsubsection{Personal Control over Anonymization}

 To establish a sense of user understanding of their personal information and privacy goals i.e. which type of the personal information should be considered most private by the service providers, we asked participants questions over control of this process and extent of privacy expected. This went in line with creating user-awareness and establishing user-control of their data security.
 
 One question explored the extent of data anonymity expected by each participant, by providing them with three separate options: 1) "all" of their data should be anonymized, 2) "some" of their data should be anonymized, 3) "none" of their data should be anonymized. The results showed a growing interest in having majority of their personal data anonymized; with "all data" option obtaining an average score of 3.875 (SD=1.17); "some data" securing an average score of 3.81 (SD=1.33); and "none data" option getting a score of 1.31 (SD=0.77).
 
 The second question ventured into the control settings for this process - if users were themselves confident in having control of these settings: "ON" or "OFF" for different sections of their personal information. This control over the anonymization process definitely falls after the user awareness of the purpose of collecting different personal data i.e. knowing back-end data requirements of IoT service providers for smooth functionality (\eg email address for account recovery). Here again, the survey results hinted at a higher preference for having control options for anonymization of data with an average score of 4.62 (SD=0.78).

\subsection{Data Processing \& Usage}
With the miniaturization of IoT modules, improvement in cloud storage, faster telecommunication technologies, and reliable networking services, majority of IoT service providers and general apps have moved towards a central data processing model i.e. the data is transferred from devices to cloud servers for processing and analysis purposes. Though this has enhanced companies' ability to provide seamless services, quicker roll-outs of updates, and global outreach, this has also raised a finite number of concerns over the security and authenticity of these remote servers. Having centralized systems can lead to bigger data breaches and costlier loss of public information. 

Additionally, many of today's services subsequently share the processed information of users with 3rd party affiliates for either gaining more insight into the data through these affiliates (to improve services) or for selling the information as marketing profiles. All of the aforementioned scenarios can have adverse affects on any user's trust in the system's reliability and credibility. Moreover, contemporary trends have hinted at a growing concern over the free services provided by different online agencies at the expense of private information of people. 

\subsubsection{Data Processing Preference \& 3rd Party Affiliates}
We asked our survey participants of their preference related to the locality of their data processing activity. With the growing usage of online services, the need for service usage data to be transferred to cloud servers is more than ever before. Many of these services need a centralized management of user information for providing split-second updates to their users. However, we wanted to observe if this process itself takes away any liberty or trust of users from these services; and whether they would appreciate local modes of data processing on their own devices rather than being part of humongous data houses. We provided the participants with two options for data processing locality: 1) data being processed locally on their device, 2) data being processed remotely on central servers. The results favoured local processing (1) of data while having an average score of 4.19 (SD=0.88) and remote processing (2) having a score of 2.19 (SD=0.95). 

An additional concern that has recently gained more traction is the subsequent usage of personal information for analytical or marketing purposes. This is occasionally powered by businesses that engage with service corporations as 3rd party affiliates, which provide either insight into user data or use it for their own marketing campaigns and sales growth. Our next question directly jumped onto this practice, as its permission is often buried under the complexities of the Privacy Agreements and impact user privacy and experience in the longer run. We are interested in unraveling whether the user experience and user trust in a service is enhanced if a service provider explicitly shares what sort of their usage information is shared with  3rd party affiliates (for any mentioned purposes). The results show that a significant amount of survey participants feel that they will be more comfortable in using a service if they were preemptively informed about the broader data sharing practices (M=4.13, SD=0.72).  

\subsubsection{Exploring Data Ownership Inclinations}
In these questions, we try to map the user behaviour in the presence of certain constraints. Though the dissemination and subsequent sale of user data to affiliates is often regarded as a controversial practice, however the possibility of user compensation for it is still an open question of research. Another question that stems from this line of queries is whether a user is even interested in maintaining ownership of their data profile throughout the entire data processing pipeline. Assuming a scenario of smart watches, this could mean having non-trivial information (\eg number of clicks, battery life usage etc.) being tagged with a user's personal information to ensure his ownership. 

We directly pose this question to the survey participants to estimate their viewpoints through a yes/no response. We ask the participants if they are willing to have their data profile tagged with personal information (\eg surname, email, IP address etc.) given three different scenarios: 1) if they receive some compensation for it, 2) if they don't receive any compensation for it, 3) if the data belongs to certain category which user is comfortable in associating himself/herself with. The survey results hint at some inclination towards favouring data ownership if there is some form of compensation related to it; hence 8 participants selecting (Yes) and 8 participants selecting (No) as their answer. For "without compensation" option, an overwhelming number of 13 participants opted for (No) to ownership tags, and 3 participants opted for (Yes). Lastly, there was a mixed response to the option (3) with 9 participants in favour of owning certain information types and 7 participants against it. 

Another question worth exploring is which private information can be used as an emblem for a user's authority over data. We present this question to the participants by providing them with a list of options that can be used as tags and fall under the broad umbrella of personal information, namely: Full name, Username, Email address, IP address, Private user key, or None of these options. The results for this don't deviate much from similar questions in these series of concepts. 8 participants feel that (None) of the tags should be used for ownership claims; whereas 7 participants feel that (Username) can serve as a good substitute and identification mechanism for ownership in data processing pipelines. (Email) and (Private key) are backed by 6 participants each, whereas (Full name) and (IP Address) are only favoured by 3 participants each owing to the easy traceability and identification through them.

\subsection{Data Profiling}
In this last section, we have tried to delve one step deeper into charting out the questions which can uncover good design practices for data privacy. Privacy concerns over personal data processing are not just related to identification, but also to the amount of information that can be garnished through intelligent analysis of this information. At the time of writing this report, Machine Learning models and similar techniques have advanced to such an extent that they can easily infer information not visible to the naked eye. And when this ability is matched with millions-to-billions of data points from IoT service users, the outcome appears to be daunting and exciting at the same time.

\subsubsection{Access to Inferred User Profiles}
In anticipation of the concern of user information being processed for inferring the user's behavioural patterns and preference profile, we present to discuss it through the perspective of user control. As many service providers rely on these data profiles to create personalized content and experience for users, we ask our survey participants if they consider having control over these inferred profiles. We try to assess their responses through the 5-point Likert scale - whether they would like to have full access of their inferred profiles (\eg shopping or music preferences etc.). Since these profiles directly impact user experience, most of the participants should willingness to have direct access to this sort of information - the average score here being 4.63 (SD=0.67).

\section{Discussion and Concluding Remarks}

The concept of IoT is very powerful given the socio-economic impacts of a connected global world. IoT has a huge potential in improving our daily lives, health outcomes, travelling safety, increased free-time at home through task automation, richer entertainment experiences, making industrial processes efficient and cheaper, and let ourselves know us better. However, these changes will come at the expense of more audio-visual sensors, cameras, thermal detectors, facial recognition and identification technology, and a handful of personally identifiable stuff. As we get to know ourselves better, and enhance our society's collective experiences of life, we also lead ourselves to a vulnerable and exposed state. With the rise in the use of IoT technologies, this work serves as a formative study following a Privacy by Design framework to elicit design requirements to navigate the tensions between benefits to the public and privacy concerns of collecting and using personal data in the IoT space.

Our survey, based on the responses from 16 participants, revealed several design considerations for privacy notifications and control in the overall process of IoT that constitutes sensing, communicating, and computing as depicted in Figure~\ref{fig:IoT_module}: 1) transparency in data collection and usage, 2) inclusive privacy efforts for wider audiences, and 3) Responsible IoT Data Management Systems.

\subsection{Transparency in Data Collection and Usage}
One critical learning from the survey responses was the gap that exists between ideal data practices for the user and what is currently performed by businesses for the service they intend to provide to the user. It is not surprising to see businesses that mandate creating a user account to use their service, which include sharing personal information to them for account authentication and password management; however, email address was the only information that our participants felt comfortable to give to the service provider. Full name, data of birth, gender, phone number, and home address, although being commonly asked during sign up depending on the service, were not considered desirable for their profile information. Additionally, the participants viewed negatively towards subsequent usage of personal information for analytical or marketing purposes and hence they would have more trust in the service if the service provider explicitly shared information on their usage involving third party affiliates. However, its permission for subsequent usage of data is often buried under the complexities of the Privacy Agreements. 

In order to match the user's data preferences and the provider's data practices, there should be a check list of information categories that the user can go through to be aware of the purpose of user data and whether certain types of data can be anonymized or opt out given informed of possible service degradation scenarios. Having options for data anonymity and opt-out data sharing was a popular opinion from our survey, as the participants preferred to be informed about the type of personal information that can be anonymized, as well as the consequences of service degradation by opting out of certain data collection. Transparency of data practices does not only benefit the user side under the growing enforcement of data minimization. It could encourage customers to see data needs as an opportunity to understand their preferences and gain the ability to make effective informed decisions for sharing their data. Taking customer consent and maintaining transparency in data collection and usage is critical to safeguarding a brand’s reputation, as well as bringing reliability and credibility of the IoT systems.

\subsection{Inclusive Privacy Efforts for Wider Audiences}
Most of the IoT applications and devices today are designed in a way that targets the use by primary users, say those who made the purchase and created a user account to control and manage the settings. This contrasts with the benefits of the connected IoT space, where subject to data collection and a recipient of service are those surrounding smart things that are not limited to primary users. We have explained a scenario of a home-entryway surveillance system that collects information on people’s comings and goings, in which subject to data collection can extend to secondary users, such as household members, and incidental users, such as guests invited to the house. From our survey findings, not to mention the secondary user groups, we learned that even incidental users would want to receive notifications for information being collected and control what data is collected. Moreover, the timings for these notifications should be a mutual agreement between the service provider and the user that varied depending on the user groups and situations. 

One approach to addressing the privacy preferences for multiple user groups is to have public notifications supported by technology. IoT devices and other systems may use the wireless connectivity to broadcast data practices to other devices nearby in so called privacy beacons~\cite{langheinrich2002privacy}. Such beacons can inform incidental users about available privacy controls~\cite{konings2013prifi}, as well as enable them to broadcast preferences to others~\cite{konings2014pripref}, such as sending a request to the primary user or the smart thing not be subject to data collection. Even though having a universal solution for everyone might not be possible, having an inclusive, informed and thorough process for developing privacy controls can help achieve the desired goals of user awareness, and mitigate some of the associated risks involved. 

\subsection{Responsible IoT Data Management Systems}

The way personal data is handled and managed by companies is one of the contributing reasons to hold privacy concerns for many people. Services share the processed information of users with third party affiliates in order to gain more insight into the data through these affiliates or worse, gain profits by selling the information as marketing profiles. We took these contemporary trends to see whether such concerns can be mitigated if there is an option for users to contribute data throughout the entire data processing pipeline, rather than share data as a trade off to use the service. We proposed an idea of data ownership in the survey if participants are willing to share their profile information, along with other data types typically collected by a service provider, under some form of compensation related to it. However, we did not observe any inclination towards this approach. They instead preferred to have authority over data and control certain categories of data that the companies can hold. In addition, they expressed their view against data being processed remotely on central servers, where the data is transferred from devices to cloud servers for processing and analysis purposes. However, with the growing usage of online services and enhanced requirements to provide seamless data communication, transferring the service usage data to cloud servers is deemed necessary.

Our survey responses encourage the ongoing discussions for designing data management systems that meet the service quality as well as trustworthiness of data mechanisms deployed in the consumer's IoT space. The participants showed more negative attitude towards current data processing models than we expected. To explore potential solutions to bridge the gap between end users and service providers, we have recently seen breakthrough in blockchain technology. It has gotten more attention and adoption in many enterprises like finance, healthcare, or supply chain~\cite{sanka2021survey}, and the IoT is not the exception. Data management using blockchain, such as proposed by ~\cite{ayoade2018decentralized} and ~\cite{dorri2017blockchain}, holds a potential promise to revolutionize the current approaches that rely on centralized systems and provide decentralized security and privacy. Also, given the lack of transparency in how user data is being shared among third party entities, blockchain-based models for data handling call for better data access permission using smart contracts and efficient trail of data access~\cite{ayoade2018decentralized}. 

While our study observed many gaps between the user's privacy goals and the data practices by service providers, our findings call for interdisciplinary research to complement different approaches at system and design levels to design privacy inclusive IoT solutions. We hope this paper helps to guide the directions for future research in the domains of privacy and IoT that benefit both end users and businesses involved.

\bibliographystyle{ACM-Reference-Format}
\bibliography{main_bib}

\end{document}